\documentclass[pdftex]{pasj01}
\Received{$\langle$reception date$\rangle$}
\Accepted{$\langle$acception date$\rangle$}
\Published{$\langle$publication date$\rangle$}
\usepackage{graphicx}

\usepackage{lineno}

\begin{document}

\title{Development of an optical photon-counting imager with a monolithic Geiger APD array}
\author{Takeshi \textsc{Nakamori}\altaffilmark{1},
		 Yuga \textsc{Ouchi}\altaffilmark{1},
		  Risa \textsc{Ogihara}\altaffilmark{1},
		 Toshio \textsc{Terasawa}\altaffilmark{2},
		 Yuhei \textsc{Kato}\altaffilmark{1},
		Shinpei \textsc{Shibata}\altaffilmark{1}
		}%
\altaffiltext{1}{Yamagata University, 1-4-12 Kojirakawa, Yamagata 990-8560, Japan }
\altaffiltext{2}{Institute of Cosmic-Ray Research, University of Tokyo, 5-1-5 Kashiwanoha, Kashiwa, Chiba, 277-8582, Japan }
\email{nakamori@sci.kj.yamagata-u.ac.jp}


\maketitle

\begin{abstract}
We have developed a sensor system based on an optical photon-counting imager with high timing resolution,
aiming for highly time-variable astronomical phenomena.
The detector is a monolithic Geiger-mode avalanche photodiode array customized in a Multi-Pixel Photon Counter
with a response time on the order of nanoseconds.
This paper evaluates the basic performance of the sensor
and confirms the gain linearity, uniformity, and low dark count.
We demonstrate the system's ability to detect the period of a flashing LED,
using a data acquisition system developed to obtain the light curve with a time bin of 100\,$\mu$s.
The Crab pulsar was observed using a 35-cm telescope without cooling,
and the equipment detected optical pulses with a period consistent with the data from the radio ephemeris.
Although improvements to the system will be necessary for more reliability,
the system has been proven to be a promising device for exploring the time-domain optical astronomy.
\end{abstract}

%\linenumbers

\section{Introduction}

A variety of high-energy astronomical phenomena, 
for example gamma-ray bursts (e.g. \cite{kum15}), fast radio bursts (e.g. \cite{pet19}), 
a neutron star merger \citep{abb17},
 X-ray binaries with Quasi Periodic Oscillations (e.g. \cite{van04}) and pulsars (e.g. \citep{eno19}),
 are known to have short time-scales variability. 
To study such objects, 
instruments with a high time resolution will undoubtedly play a crucial role.
In optical astronomy, very short time-scale phenomena, 
typically shorter than one second, still remain relatively unexplored compared with studies at other wavelengths
such as radio, X-ray and gamma rays.
Generally, due to a power-law spectrum, 
number of photons from non-thermal radiation are even higher in optical wavelengths than in X-rays or gamma rays.
Since photon statistics is important for detecting time variability,
optical observations may have an advantage over higher energies, even though often contaminated with thermal emission.
The variability time scale $\tau$ restricts the size of emission  region as
$c \tau/\Gamma = 3~(\tau/1\,{\rm ms})(\Gamma /100)^{-1}$\,km,
where $c$ is the speed of light and $\Gamma $ is a bulk Lorentz factor of emitters. 
Highly time-resolved observation with plenty photons could be a powerful tool for probing very small scale structures.

A well-established optical instrument is the Charge Coupled Device (CCD),
which integrates photons over a frame exposure.
The time resolution of a CCD is limited mainly by its read out time, 
typically on the order of a second for a large CCD.
Recently Complementary Metal-Oxide-Semiconductor (CMOS) imaging sensors have been developed for astronomy
and are already in use.
For example, the Tomo-e GOZEN camera mounted on the Kiso-Schmidt telescope has demonstrated a wide field-of-view (FoV) and fast readout, namely 2 frames/s in full frame mode and $\sim$500 frames/s in partial readout mode \citep{sako18}. 

Using photon-counting devices is another way to achieve fine time resolution,
since such detectors have an extremely fast response time. 
A photomultiplier tube (PMT) has been the device of choice, 
offering fast response and a large internal gain, typically $10^{6}$ to $10^8$.
High Speed Photometer \citep{Bless+99} once onboard the Hubble Space Telescope utilized a PMT
to obtain the light curve of the Crab pulsar at visible and ultraviolet wavelengths 
with $\sim 20\,\mu$s resolution \citep{Per+93}.

ARCONS \citep{maz+13} has developed microwave kinetic inductance detectors for astronomy
that enables photon counting and spectroscopy from visible to infrared wavelengths.
Although ARCONS requires cooling by a cryostat,
its excellent performance was demonstrated by the significant detection of the enhancement of the Crab pulse 
accompanying the giant radio pulse \citep{str+13}.

Silicon semiconductor devices have also been studied intensively, in particular,
the family of Geiger-mode single-photon avalanche photodiodes (SPAD).
OPTIMA \citep{str+01} discovered an X-ray and optical correlation from black hole candidate \citep{kan01},
and magnetar flares \citep{ste+08}.
Optical pulsation from milisecond pulsars were detected by SiFAP \citep{amb+17} and Aqueye+ \citep{zam+19}.

The Multi-Pixel Photon Counter (MPPC), also commonly known as a silicon photomultiplier, 
is a semiconductor photo-sensor 
that consists of many avalanche photodiodes ("cells"  "microcells") in a two-dimensional array \citep{yam06}. 
Each cell works in the Geiger mode 
in order to employ an internal gain of $\sim 10^6$,
which allows even a single optical photon to be detected.
All the cells are connected in parallel 
so that net output signal is proportional to 
the number of photons detected at the same time in each cell. 
Its fast response with a time jitter typically of 100\,ps 
is also suited for precise measurements of photon arrival times.  
However, MPPCs also generates "dark counts,"
which are spurious pulses that cannot be distinguished from real signals due to photon detections. 
The dark count rate depends on temperature, the size of the active area, and the operating voltage.
A typical dark count rate is several hundred kHz, 
so using MPPCs for astronomical observation is challenging, especially for faint sources. 
However, it should be noted that the ground-based imaging atmospheric Cherenkov telescopes
are exceptions. They actively employ MPPCs or SiPMs for their cameras.
FACT \citep{and+13} has been in operation for several years, and
ASTRI-Horn \citep{lom+20} was developed within the framework of the Cherenkov Telescope Array \citep{ach+13}.
In these systems, coincidence between many pixels and adequate trigger setting suppress the effects of dark counts.
In a similar way, applying such a coincident technique,
 \citet{li+19} developed an optical observation system with MPPCs.
  
This paper presents an optical photon-counting observation system 
with uniquely customized MPPCs and significantly reduced dark count rate.
Such a simple and compact system could be easy to handle, carry and replicate, 
and applicable to a variety of observational targets and strategies.
This paper is organized as follows.
Section 2 reports the structure and performance of the sensor system.
Section 3 describes the data acquisition system and the instrument accuracy.
Section 4 explains our observation procedures for celestial objects.
Section 5 presents observational results of the Crab pulsar.
The conclusions at the end discusses the overall performance of this prototypical system and future applications.

\section{Detector and characteristics}
\begin{figure}
\centering
\includegraphics[width=7cm]{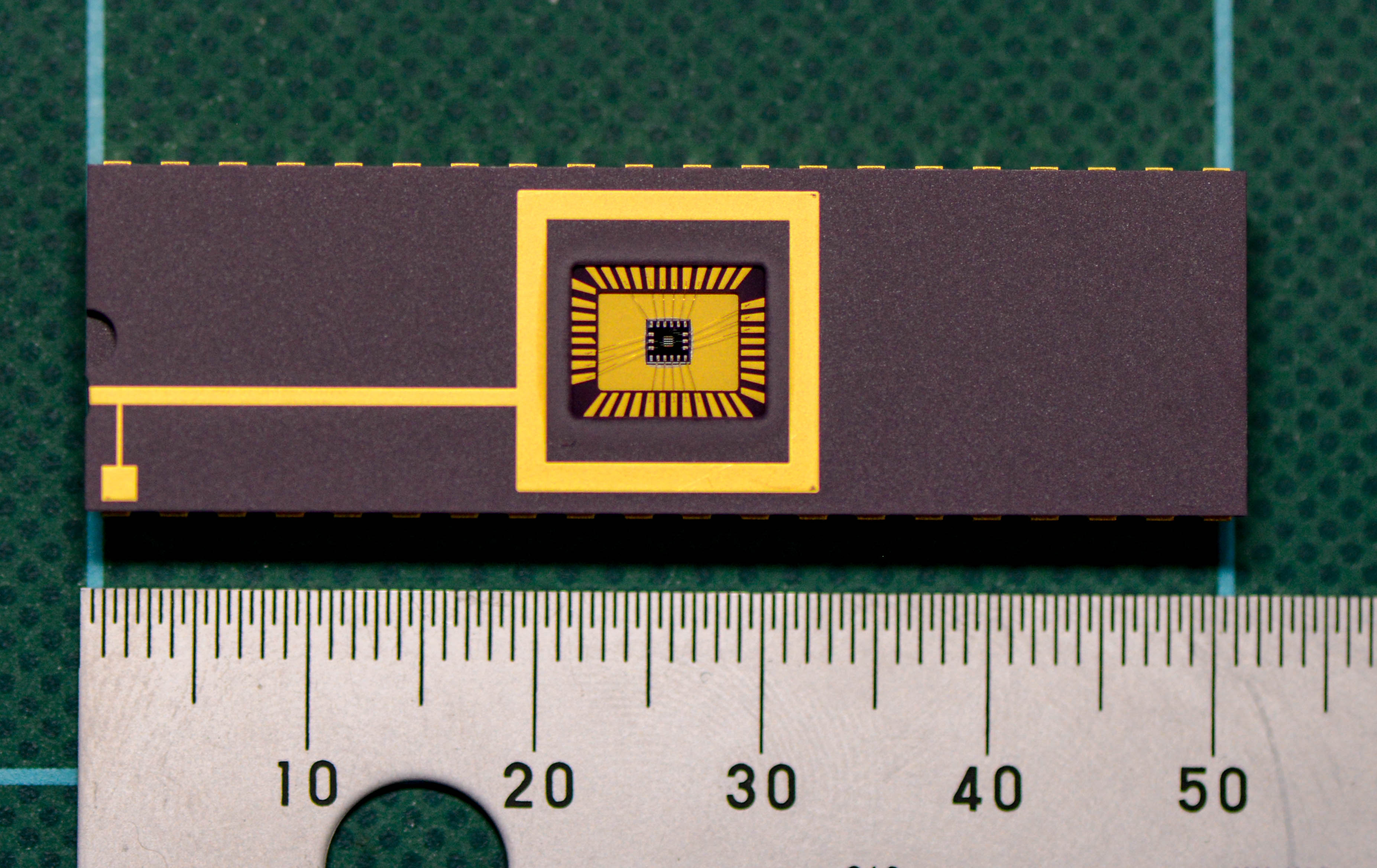}
\caption{A photo of the customized MPPC or GAPD array. 
Common cathode and 16 anodes are connected to the pins.}
\label{fig:gapd}
\end{figure}

Figure~\ref{fig:gapd} shows a customized MPPC as a dual-in-line package (Hamamatsu, S13361$-$9088) . 
The sensing area consists of a $4\times 4$ array of $100\times 100\,\mu$m$^2$ cells providing a total of 16 channels.
Unlike commonly available products,  
the anodes of these individual cells are not connected to each other,
and every cells work as independent Geiger avalanche photodiode (GAPD).
This structure gives the customized MPPC (hereafter GAPD array) a sensitivity to the arrival position of photons
and can be operated as an imager.
The advantage of a monolithic GAPD over an assembly of individual APDs is that
its channel characteristics are naturally uniform.
Typical photon detection efficiency (PDE) curves provided by the manufacturer \footnote{hamamatsu.com}
give an operating wavelength of $300-900$\,nm, with a peak about 450\,nm.
These PDE values include the filling factor for the sensitive area,
so absolute quantum efficiencies can be estimated as high as 70\% around the peak.
Since the sensing area for each channel is much smaller than that of an MPPC, 
a lower dark count rate can be expected.
For example, the dark count rate should be more than two orders of magnitude less than 
that for a commercial $1.3\times 1.3$\,mm$^2$ MPPC.
The low dark count rate is one of the most important aspects 
in the application of the GAPD array to astronomical observations.

\subsection{Pulse shape}
\begin{figure}[h]
\centering
\includegraphics[width=7cm]{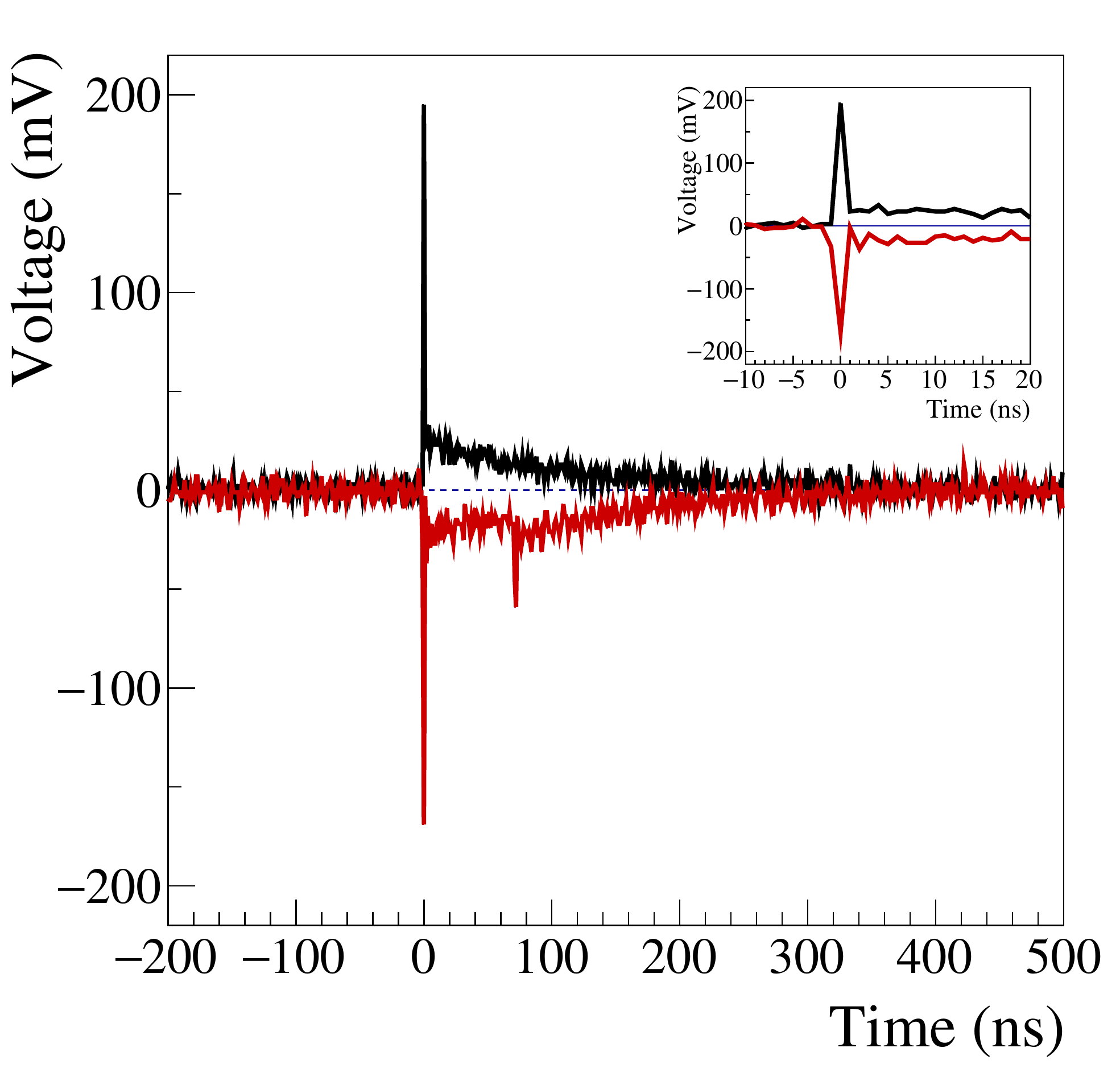}
\caption{Examples of the pulse shape of a single pulse (black line, positive) .
	A second pulse, represented on an inverted scale, is accompanied by an after-pulse (red line, negative).
	The inset panel shows a close-up view around the main pulses. 
	The sampling interval is 1\,ns.}
\label{fig:pulse}
\end{figure}

We fed the output signal from a channel to two $\times 10$ fast amplifiers (Philips, 775) connected in series, 
achieving a gain of $\sim 100$, 
and then to an oscilloscope with a bandwidth of 1\,GHz (Tektronix, MSO4014B-L). 
Figure~\ref{fig:pulse} shows examples of pulse shapes with and without an after-pulse
that appears at $\sim 80$\,ns later than the main pulse.
During this measurement the detector was placed in a thermal chamber at $25{}^\circ$C,
and a bias voltage of $V_{\rm OV}=3.0$\,V was applied (see also the following subsection).
A sharp pulse shorter than 5\,ns duration is clearly seen,
followed by a tail of the slow component. 
The smaller area of each channel is responsible for the short initial pulse width,
due to the smaller capacitance and hence shorter time constant.
Simply by using a leading-edge discriminator, 
timing precision for the pulse detection is expected to be on the order of nanoseconds. 
After-pulses could cause an overestimate of the number of pulses.
Since the wave heights are significantly different,
however, it can be sufficiently well discriminated if the threshold has been chosen properly.

\subsection{Basic characteristics}
\begin{figure}
\centering
\includegraphics[width=7cm]{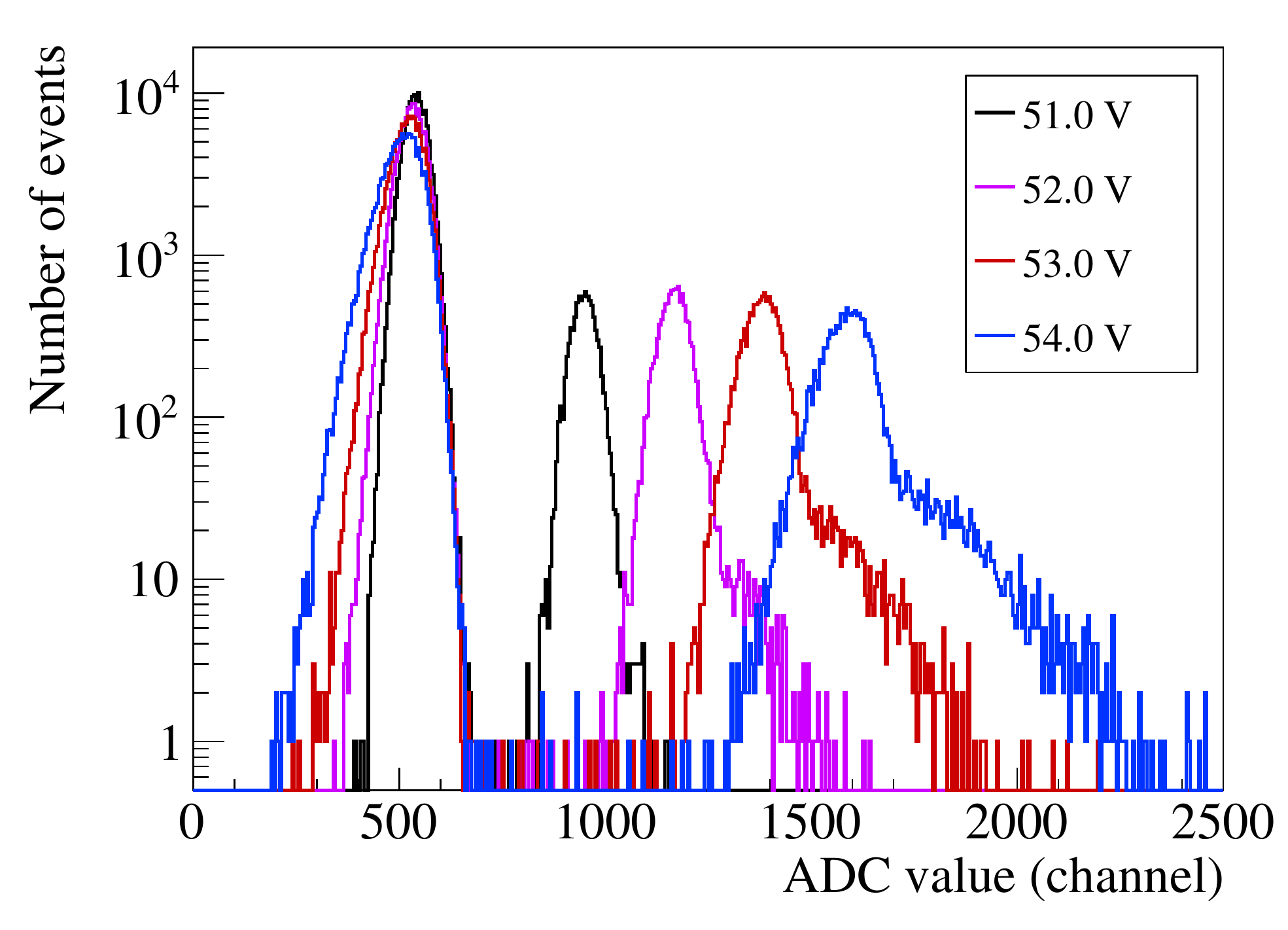}
\caption{An example of 1\,photoelectron spectra obtained by the channel 1 with different bias voltages.
 These measurements are performed at $-20^\circ$C.}
\label{fig:spec1pe}
\end{figure}

The GAPD array was illuminated by a light emitting diode (LED)
and the data acquisition was triggered by a synchronized pulse.
The GAPD signal was amplified and recorded by a charge sensitive ADC (Hoshin, V005)
with an integration period of 200\,ns to include the slow component.
Figure~\ref{fig:spec1pe} shows  the charge spectra for various bias voltages.
The single photon peak and the pedestal are clearly separated
and no other peaks, corresponding to two or more photons, are seen (as expected). 
This is because each cell detects only one photon at a time.
Since the integration time is long enough to contain the after pulses,
the single photon peak has a tail to the right. 
Apparently the fraction of the tail component increases with gain.
This can be naturally understood by the known fact that
 the greater gain increases the probability of after-pulsing.

\begin{figure}
\centering
\includegraphics[width=7.5cm]{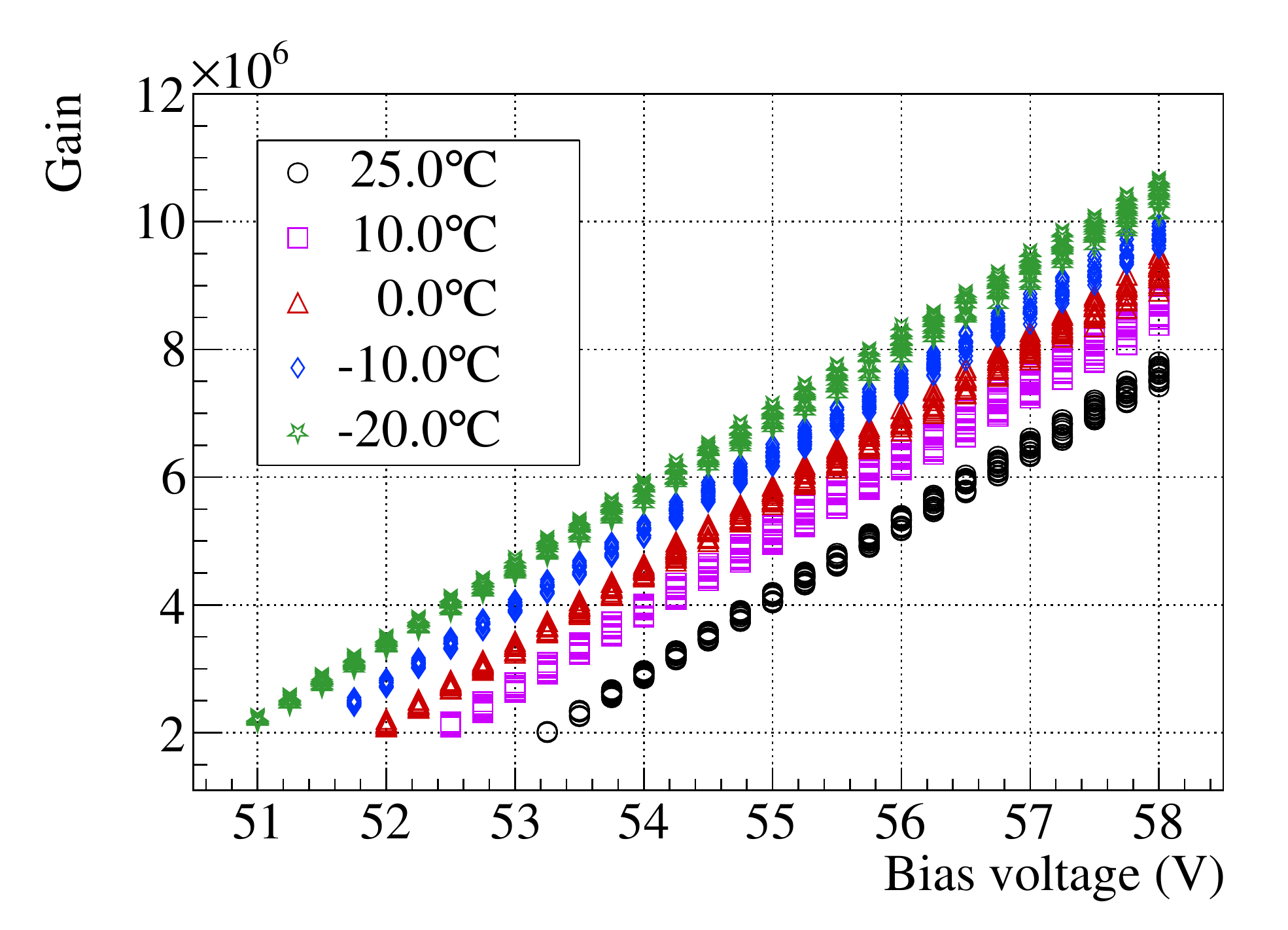}
\caption{The gain curves as a function of the bias voltage,
for all 16 channels and various temperatures.}
\label{fig:gain_linear}
\end{figure}

\begin{figure}
\centering
\includegraphics[width=9cm]{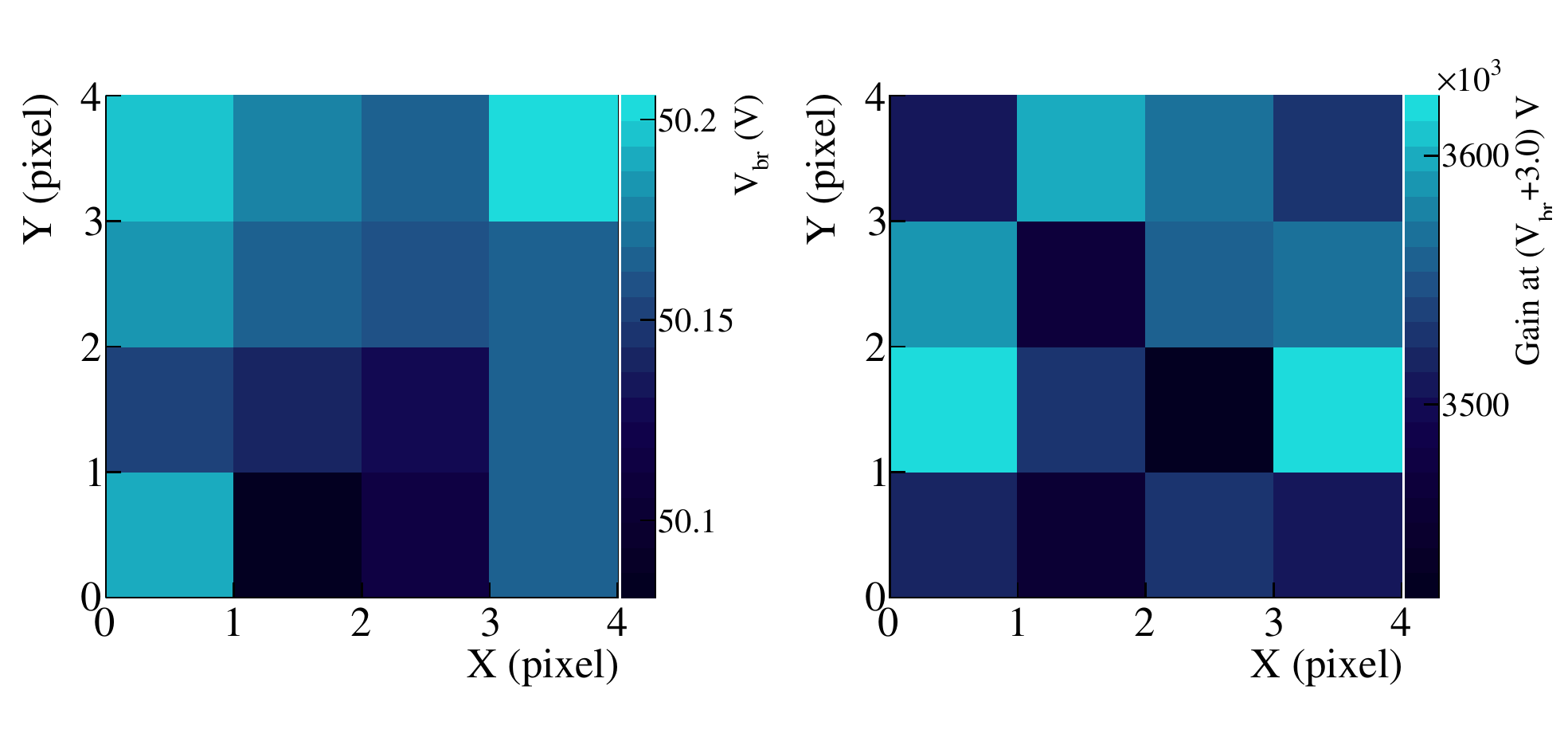}
\caption{Maps of the breakdown voltages (left) and gain at operating voltage (right). 
	$X$ and $Y$ positions correspond to those on the sensor.
	Both data are derived from the results at $0^\circ$C.  }
\label{fig:uniformity}
\end{figure}

Figure~\ref{fig:gain_linear} shows the gain curves as a function of the bias voltage,
for all 16 channels and various temperatures.
Commonly known features of MPPCs are clear: 
the gains are proportional to the over-voltage;
lowering the temperature decreases the breakdown voltage $V_{\rm br}$,
and this leads to higher gain at the same bias voltage.
The variation across the array of the breakdown voltages and gains at the operating voltage $V_{\rm op} =V_{\rm br} +3.0$\,V 
are shown in figure~\ref{fig:uniformity}. 
The dispersion is less than 1\% and 3\% in the $V_{\rm br}$ and the gain, respectively.

\begin{figure}
\centering
\includegraphics[width=7cm]{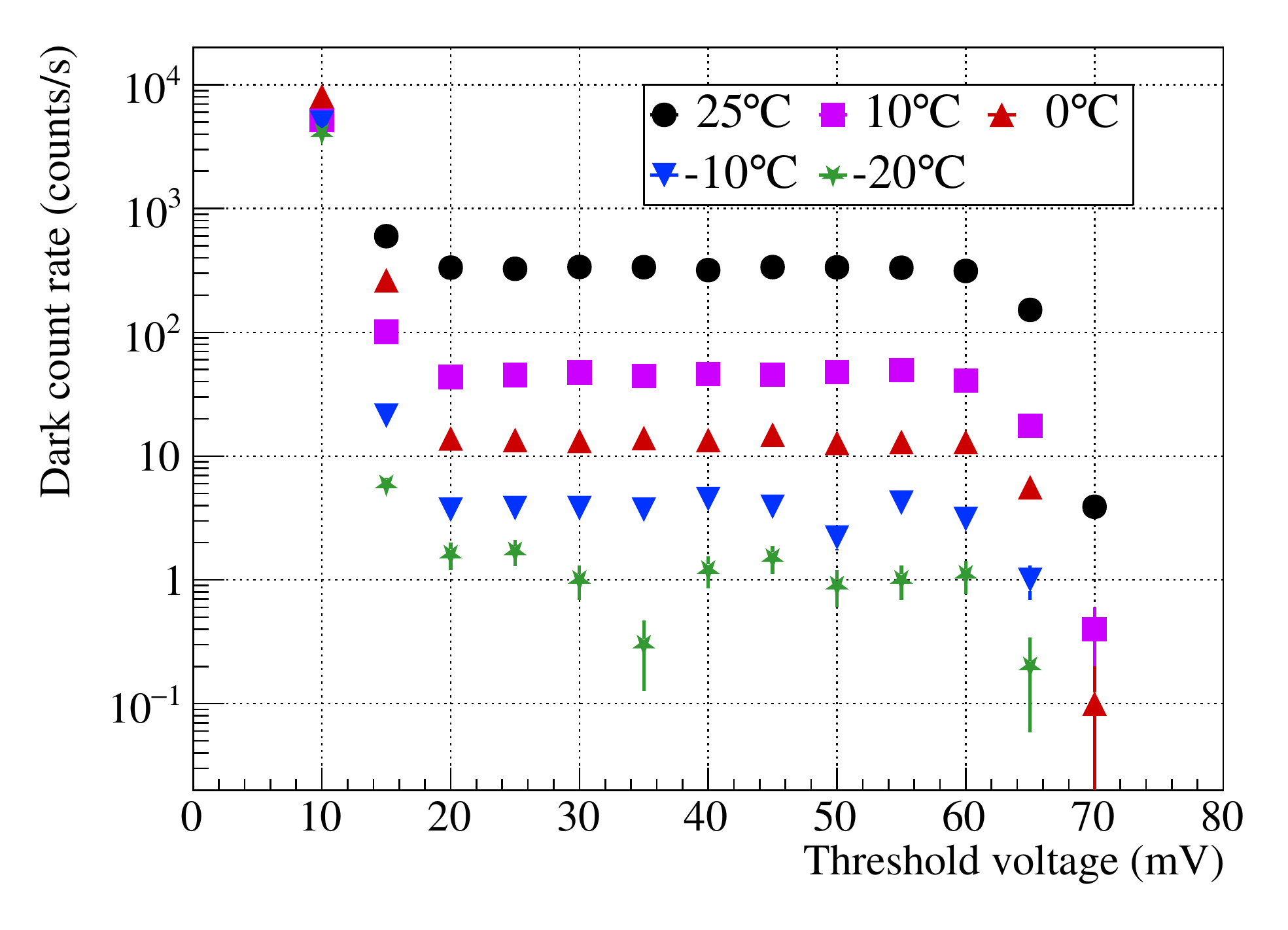}
\caption{An example of dark count rates as a function of the threshold voltage.
The circle, square, upward- and downward-pointing triangles and stars 
correspond to the data taken at $25, 10, 0, -10$ and $-20^\circ$C, respectively.
}
\label{fig:darkscan}
\end{figure}

\begin{figure}
\centering
\includegraphics[width=7cm]{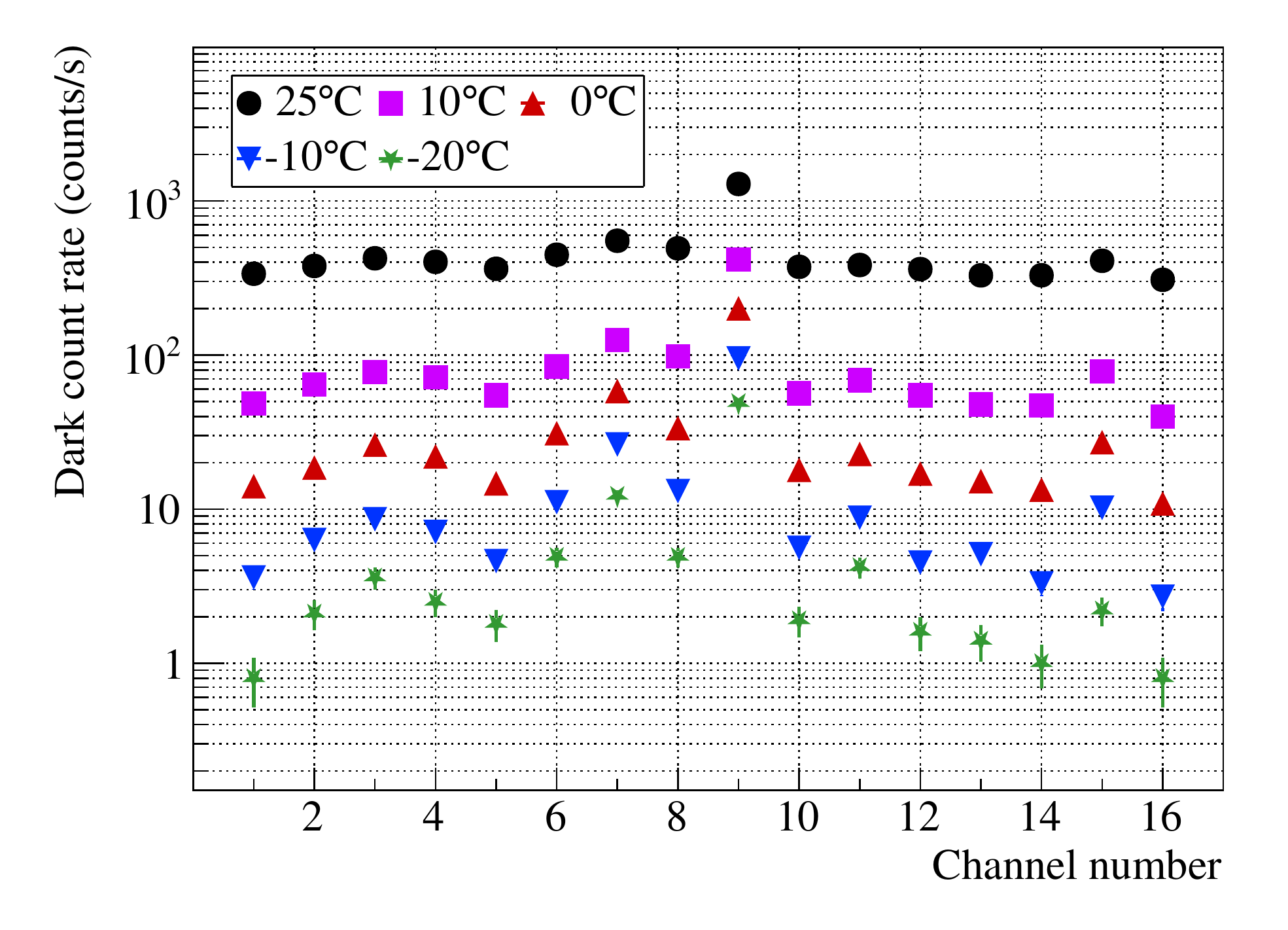}
\caption{Dark count rates for all the 16 channels at various temperatures.
The circle, square, upward- and downward-pointing triangles and stars  
correspond to the data taken at $25, 10, 0, -10$ and $-20^\circ$C, respectively.
}
\label{fig:darkscanall}
\end{figure}

Another important characteristics is the dark count rate.
Bias voltage is set at $V_{\rm op}$ for all the following measurements performed.
Figure~\ref{fig:darkscan} shows the dark count rate as a function of the threshold voltage on temperature. 
At each threshold, the rates calculated using time bins of 10\,s are plotted.
Below 15\,mV, circuit noise is dominant.
There is a stable plateau above 20\,mV;
above 60\,mV, the rate drops sharply as the threshold exceeds the 1\,photoelectron. pulse height.
 Each plateau level indicates a dark count rate equivalent to the 0.5\,photoelectron threshold.
Note that each channel consists of a single cell and hence multi-photon components are not observed.
More noteworthy is the low dark count rate, which is one-hundredth that of a commercial MPPC,
as expected from the element area ratio.
Figure~\ref{fig:darkscanall} shows the dark count rate at the plateau for all 16 channels. 
The dark count rates of all 16 channels, with the exception of channel 9, are concentrated around 400\,counts/s at room temperature.
With lower temperature, the rates drop for every channels, 
though the dropping ratio is slightly different for each channel.
The characteristics of each channel are likely due to multiple factors
and are not easy to identify, so that we do not discuss them further in this paper.
Two possible causes are a difference of intermediate levels in the band gap
and local concentration of the electric field due to poor pattern formation.
It is important to observation as to whether the dark count rate is higher than the night sky background.
This point is discussed later.

Finally, the optical crosstalk (OC) probability was evaluated.
A fired cell sometimes emits photons those might trigger the Geiger discharge in neighboring cells.
This is called the OC that generates a correlating signal without incident photons.
In general, thicker enclosure lowers the OCT probability (e.g.,\cite{asa18})
and the GAPD array used in this work has a silicone resin of $0.7\pm 0.2$ mm thick.

The OC pulses triggered by the dark count were observed
with the bias voltage of $V_{\rm op}$ and the temperature of 0$^\circ$C.
The waveforms of each cell is recorded at 1-G\,samples/s with $>500$\,MHz bandwidth
by a VME-based waveform digitizer module (V1742; CAEN).
The offline pulse search analysis was performed 
and the timing of the pulses was identified for each cell.
Pulses from two different cell within a time window of 20\,ns are selected \citep{otte17}
and the earlier pulse was considered to initiate the later pulse due to the OC.
%These measurements and analyses for each cell can only be performed with the GAPD array, not with commercial MPPCs.

\begin{figure}
\centering
\includegraphics[width=7cm]{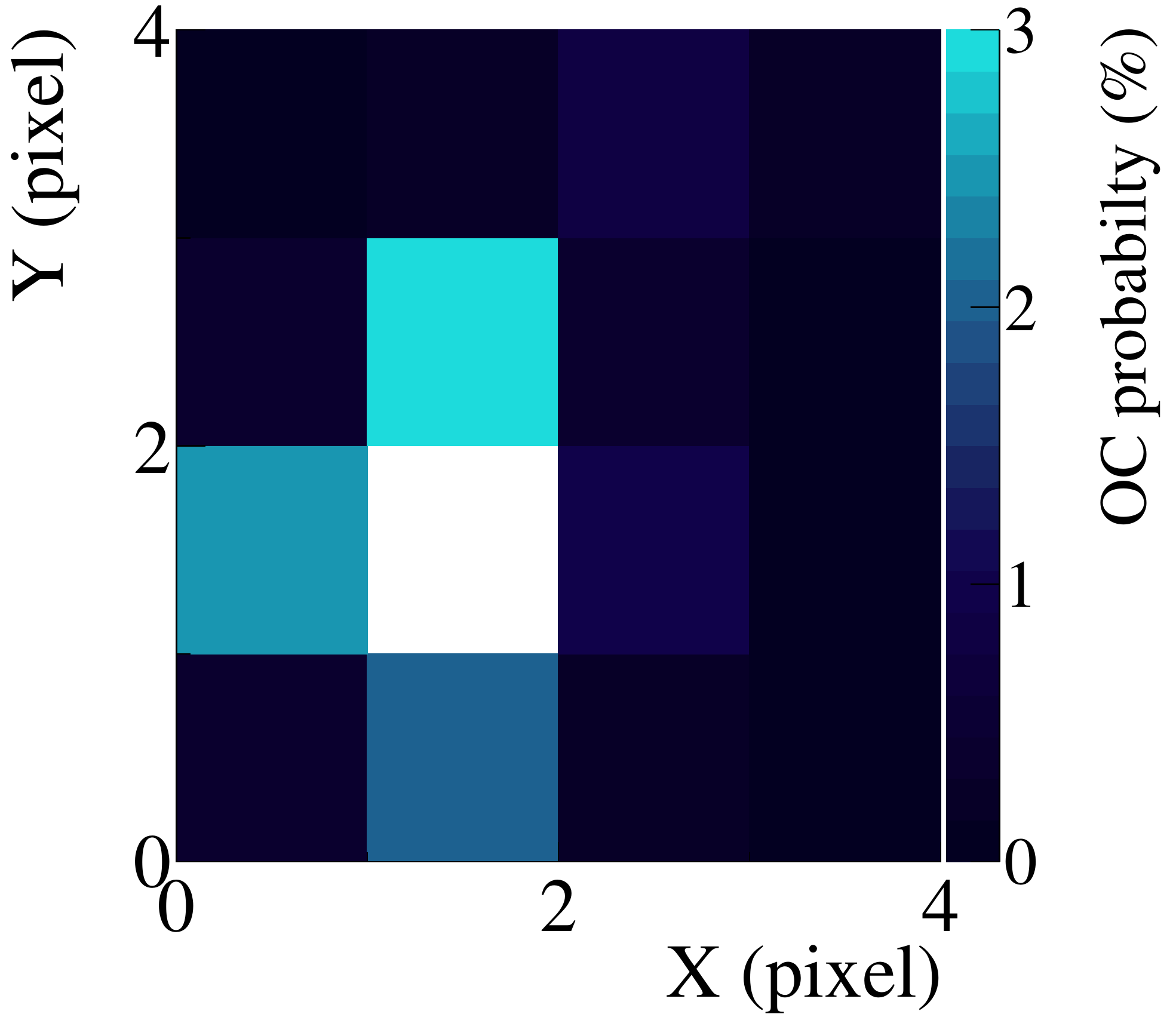}
\caption{ A map of the OC probabilities triggered by channel 6 indicated by a blank square.
$X$ and $Y$ positions correspond to those on the sensor.
}
\label{fig:oct}
\end{figure}

Figure~\ref{fig:oct} shows a map of the OC probabilities triggered by channel 6 located at the position indicated by a blank square.
For the clarity, events where a channel 6 signal is considered as an OC pulse triggered by other cells are removed.
As a result, the number of survived events are $\sim 5\times 10^4$.
Apparently the closer cells tend to be higher OC probability as expected,
and the highest probability of 2.8\% was observed at channel 2, which is at the left of channel 6. 
Despite the geometrical symmetry of the package structure, the asymmetric OC distribution was observed.
The reason for this asymmetry is not clear, but possibly due to slight differences in Geiger discharge probabilities among channels.

%The dark count rates shown in Figure~\ref{fig:darkscanall} may contain OCT pulses.
%However, the rates are so low that chance coincidence probability of the dark counts between the cells under analysis
%is less than $2\times 10^{-5}$ for channel 9 with the highest rate.

\section{Data acquisition and instrument accuracy}
\begin{figure}
\centering
\includegraphics[width=9cm]{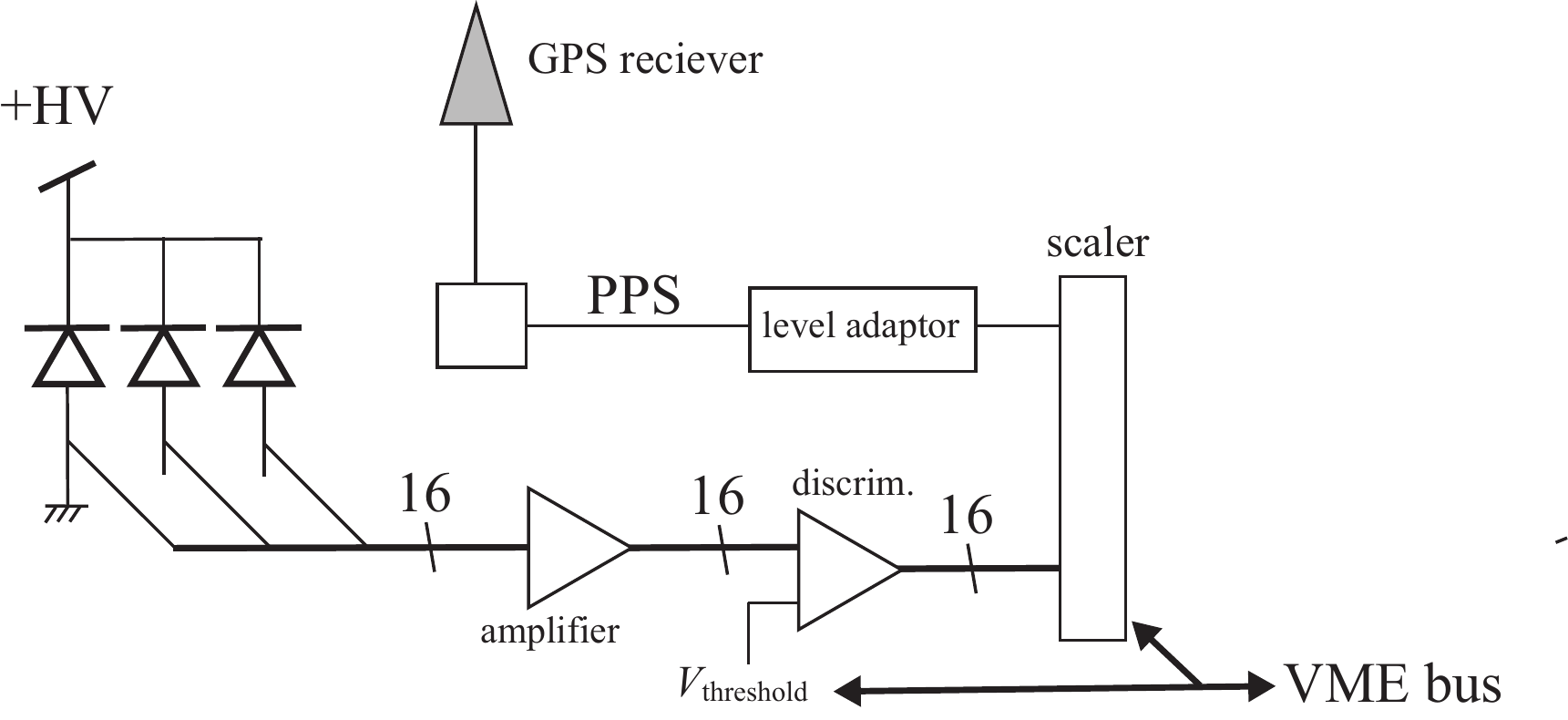}
\caption{Setup of the data acquisition system. }
\label{fig:daq}
\end{figure}

Figure~\ref{fig:daq} is a schematic view of the data acquisition system 
for recording light curves without dead time for any of the pixels.
The output signal from the sensor is fed into inverting amplifier circuit
which employs fast AD8099 amplifiers.
When a signal equivalent to 1 p.e. is generated,
a 16 channel discriminator (CAEN, V895) generates digital pulses.
The threshold level can be set independently for each channel
so that the system is flexible to pulse height variations
caused mainly by the gain and offset variation of the fast amplifier circuits.
Finally a scaler (CAEN, V830) counts the number of pulses in every 100\,$\mu$s interval for each channel.
The scaler is triggered by its internal clock and data are stored in an onboard buffer,
independent of data transfer processes.
This function enables our measurements to be completely free of dead time.
Pulse-per-second (PPS) signals from a GPS receiver are also connected to the scaler. 
The GPS pulses tag a time stamp on the recorded light curve bins once every second,
and the absolute time can be calculated using these tags and the clock of the DAQ computer synchronized by ntpd.

\begin{figure}
\centering
\includegraphics[width=8cm]{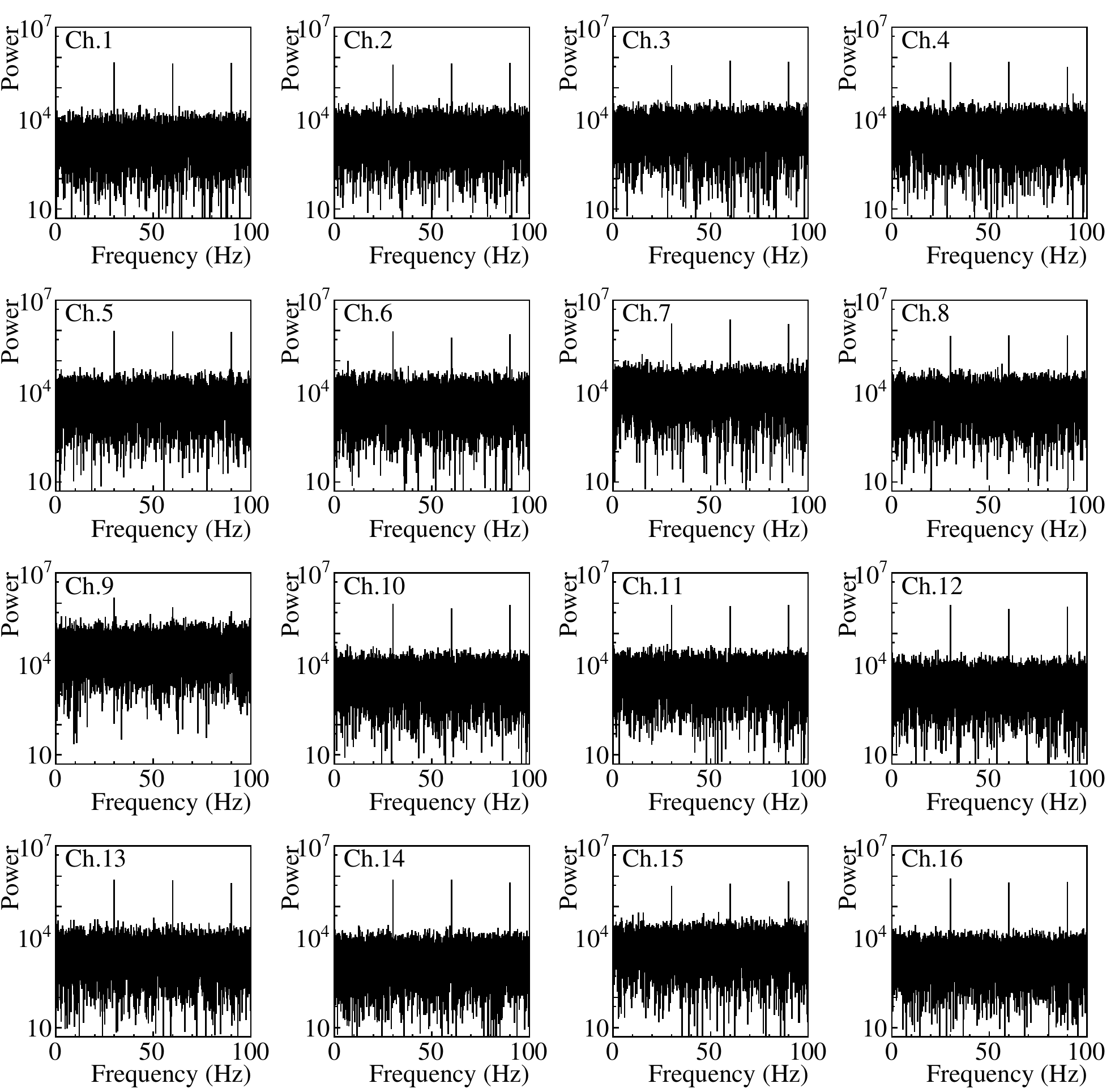}
\caption{Power spectra for the repeated 30-Hz LED flashing data for each pixel.}
\label{fig:power}
\end{figure}

To demonstrate the capability of period detection with high time resolution, 
we irradiated the GAPD array repeating LED flashes and recorded the light curves.
The frequency of the flash was 30\,Hz triggered by a function generator
with a pulse width of 100\,ns.
This measurement was performed at $0^\circ$C.
The exposure time is 60 seconds and all the pixels are measured simultaneously.
Figure~\ref{fig:power} shows the Fourier power spectra produced from the light curves for each pixel.
For all pixels, a clear peak at $30.00\pm0.02$\,Hz and the second and third harmonic peaks are clearly seen.
The first peak frequencies are consistent with that of the light source, 
and the spectra are consistent with that to be expected from the $\delta$-function-like illumination light curve.
The power spectrum of channel 9 has the worst signal-to-noise ratio,
since the dark count rate of this channel is the highest among the 16 channels.

\begin{figure}
\centering
\includegraphics[width=8cm]{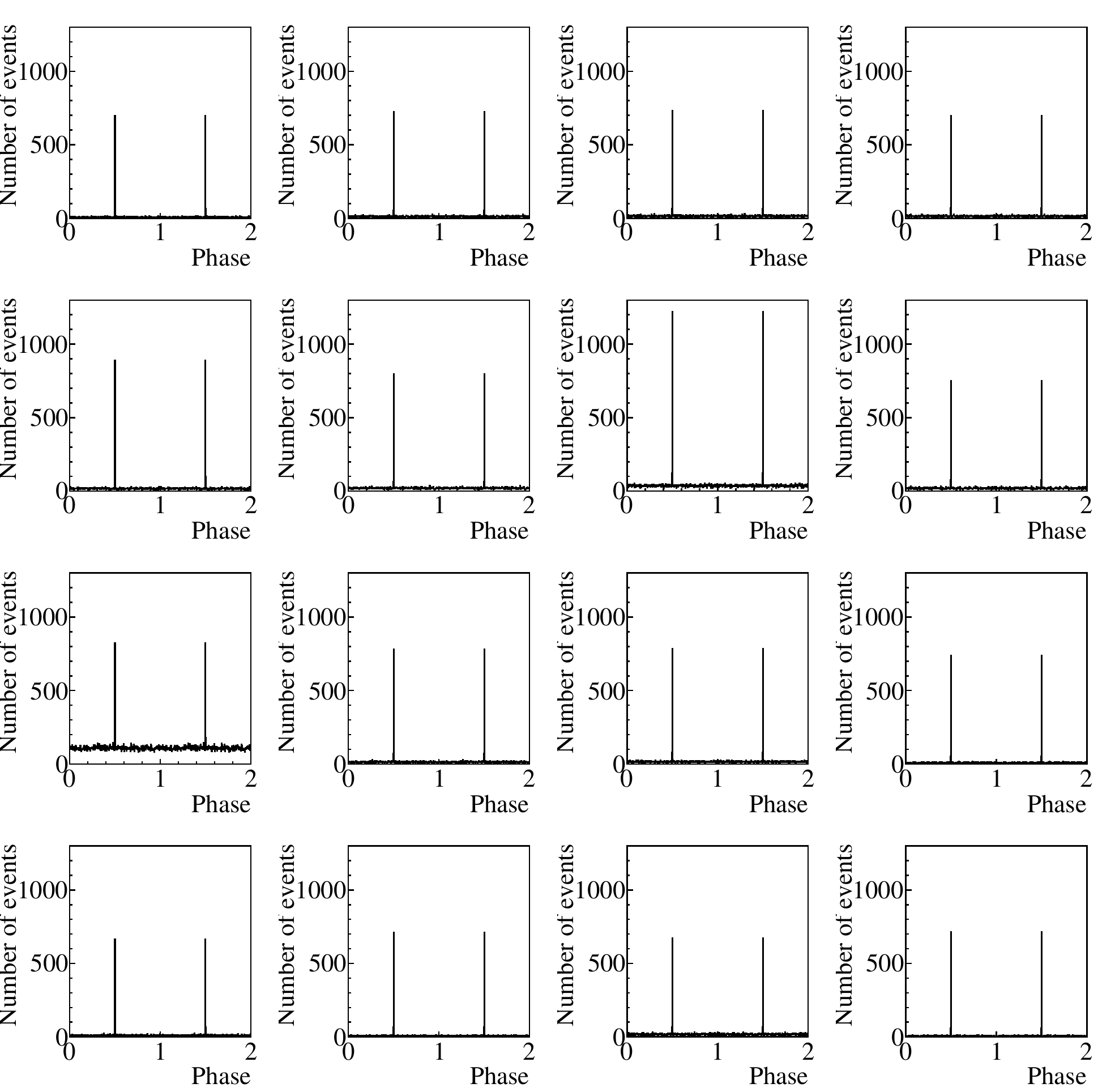}
\caption{Folded light curves of the repeated 30-Hz LED flashing data for each pixel.
Note that the two cycles of the periodic light curves are shown.
}
\label{fig:ledfold}
\end{figure}

Figure~\ref{fig:ledfold} shows the folded light curves with a frequency of 29.9997~Hz
that is the best value derived from the epoch folding technique.
The number of bins for a cycle is 333, which corresponds to $\sim 100\,\mu$s.
The statistical error is estimated as $\pm 0.0003$\,Hz following \citet{lar96}.
which is consistent with the light source (Agilent, 33120A) frequency of 30~Hz.
The detection time stamp of the LED flash is very well contained 
in a single time bin of $100\,\mu$s for all the pixels as expected.
The flat offset component in each light curve corresponds to the dark count,
which occurs randomly independent of the flash.
One can also find that channel 9 has the highest dark count rate from this figure. 
We also confirmed the capability of higher frequency detection
using the LED, by raising the flashing frequency of 1\,kHz and more.
For example, $999.9980\pm 0.012_{\rm stat}$~Hz was obtained
from 1-minute observation of the 1~kHz light source.

From these demonstrations, 
we conclude that our detector and system are able to observe light curves with $100\,\mu$s time resolution.

\section{Observations}

We decided on the Crab pulsar as the first target 
to demonstrate the performance of our time-resolving observation system.
This was not only because the Crab pulsar is the brightest pulsar at visible wavelength and its optical light curve is well known (e.g., \cite{zan14}),
but also because its ephemeris is continuously provided by radio facility \citep{Lyn+93} and monitored by various observatories from radio frequencies to gamma rays.

\subsection{Instruments and observational setup}
\label{sec:obs}
All the observations reported in this paper were conducted at the Yamagata Astronomical Observatory,
which is located on a roof of Yamagata University.
The Cartesian coordinates of the observatory is $(X, Y, Z) = (-3861744, 3200488, 3927194)$~m.
It is in the middle of the city of Yamagata and mainly for amateur use. 
Because of this environment,
the night sky background is expected to be considerably brighter than that of other observatories. 
However, the light pollution from the city is not expected to be a problem in this work.

Observations were made with a 35-cm diameter telescope
which is commercially available Advanced Coma-Free optics (Meade, F8ACF)
and is mounted on an equatorial mount (Takahashi, EM-400 FG-Temma2Z).
We also employ an automatic guiding system (Laccerta, M-Gen) 
to improve the tracking accuracy.
The focal length of the telescope is 2845\,mm so that the FoV of the sensor is 
$\sim  28.8^{\prime\prime}\times 28.8^{\prime\prime}$ or $7.2^{\prime\prime}\times 7.2^{\prime\prime}$/channel.
The equatorial mount is controlled by software (Stella Navigator version 10, Astro Arts).
Only a manual focussing system is available by turning a built-in knob by hand.
The point spread function (PSF) is roughly estimated as $\sim40\,\mu$m or $\sim 3^{\prime\prime}$ (FWHM) in advance 
by taking images of stars with a digital camera.

\begin{figure}
\centering
\includegraphics[width=7cm]{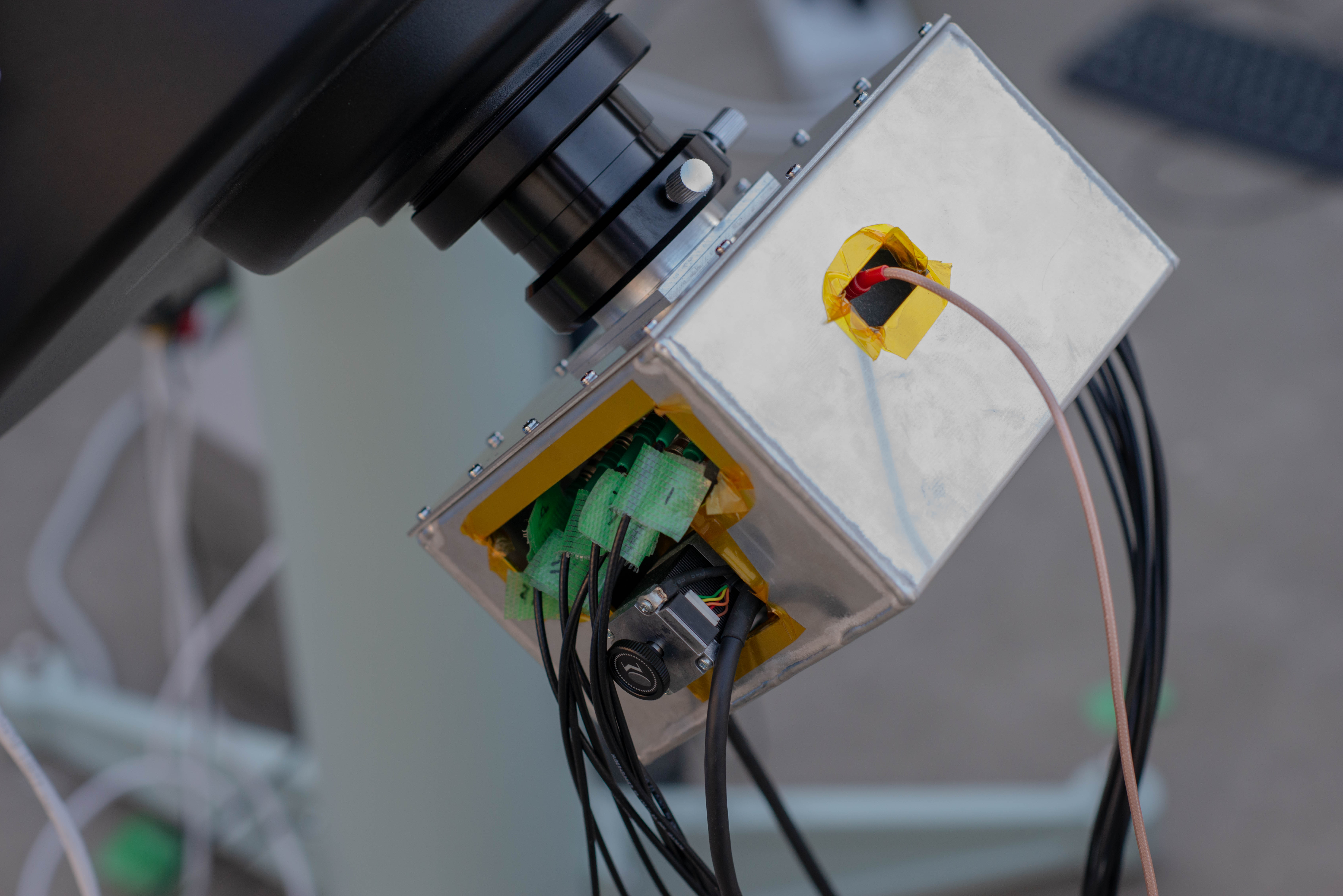}
\caption{A photo of the imaging box connected to the telescope.
}
\label{fig:jig}
\end{figure}

We fabricated an imaging box consisting of a sensor jig 
fixed on an XY stage with an accuracy of $1\,\mu$m (Sigma Kouki, TAMM40-10C(XY)). 
The XY stage is remotely controlled via serial communication.
We mechanically connected the imager box to the telescope as shown in figure~\ref{fig:jig}.
By moving the stages at the focal plane of telescopes, 
we could effectively cover a wider FoV by mosaic imaging
even with the very small sensitive area of the GAPD array.
%This function is quite useful not only for the focussing but also for pointing at targets
%played a crucial role in evaluating our primitive system.
%Since our system does not employ any other imaging subsystem,
%the mosaic image is the only means for the optical axis alignment.

\subsection{Procedure}
%Figure \ref{fig:seq} shows an observation procedure 

Before starting observations, we measured the dark count rate for each channel with the telescope lid closed,
and determined the corresponding threshold levels for each channel.
The observatory is exposed to the open air and 
the temperature was not controlled.
The sensor box has no mechanical cooling system 
since no devices in the box generate significant heat. 
The air temperature during the observations was at least $5\pm5{}^\circ$C and stable
so that the threshold levels were not changed until the end.
And the large signal-to-noise ratio for every pulse enabled us 
to determine appropriate threshold voltages with wide margins. 
Therefore the system is tolerant against any gain variation due to temperature fluctuations.

The detailed of the operating procedure up to focussing on a star image are descried in Appendix~\ref{app:focus}.
After focussing, data for the flat-field correction and for the normalization of detection efficiency
was obtained as follows;
Every channel of the GAPD array was exposed to the same sky region by repositioning the XY stage.
The correction coefficients were calculated 
so that the counts are the same after subtracting the corresponding dark counts for each channel.
The exposure time was determined to have a statistical error of 1\% or less.
From this measurement the dark-sky background rate was as high as $\sim 2$\,kcounts/s,
which is significantly higher than the dark count rate at the temperature of $5^\circ$C or so, 
even when compared to the channel 9 with the exceptionally high dark count rate.

The Crab pulsar region was observed from 11:45 to 13:35, 2019 March 9 (UTC),
with elevation angles of $63^\circ.4$ to $44^\circ.0$.
First, a mosaic of $5\times 5$ images was taken, each with a 1-second exposures 
to confirm telescope alignment to the approximate position of the Crab pulsar. 
We observed the region around the pulsar position performing $4\times 4$\,mosaic acquisitions. 
Each frame of the mosaic has a duration of 1\,minute. 
Completing all the frames of the mosaic takes approximately $20-25$\,minutes. 
The mosaic acquisitions were repeated seven times. 
As the Crab pulsar falls inside a single frame, the net effective exposure was 7\,minutes.
%The light curves were then measured for one minute in each $4\times 4$ region around the pulsar in sequence.
%,allowing for telescope slew.
%Net effective exposure to the Crab pulsar was 7 minutes as a result.
%Sometimes thin cloud that was identified even by eyes passed across the line of sight.
The sky condition was rather fine and stable.

\section{Analysis and results}
\subsection{Image}
\begin{figure}
\centering
\includegraphics[width=7cm]{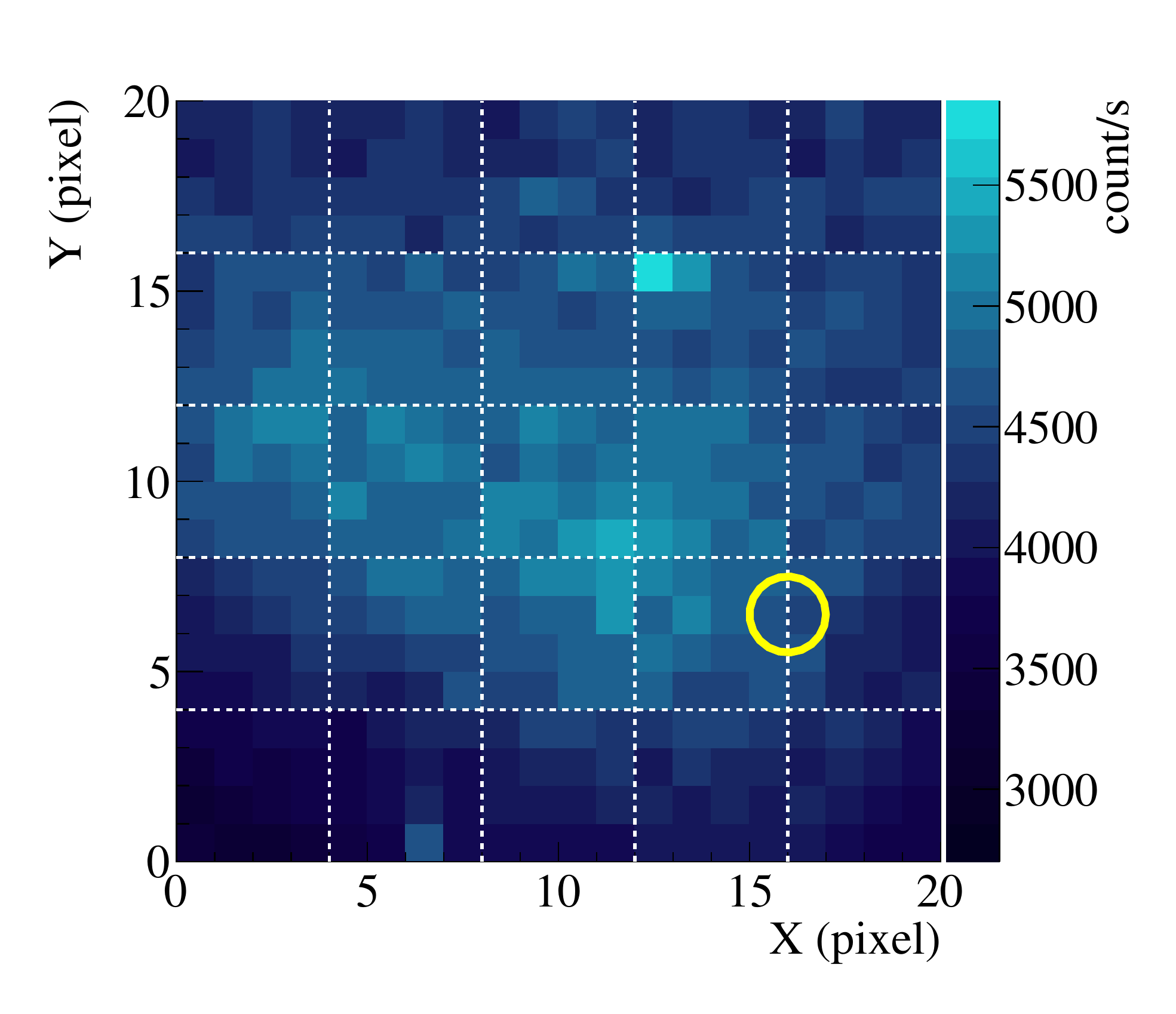}
\caption{Count map of the Crab pulsar region, composed of $5\times 5$ images
(corresponding to $2.4^\prime\times 2.4^\prime$).
 The dark counts are subtracted and the flat correction is applied.
 The position of the Crab pulsar is indicated by a circle.
 Note that the target position slightly shifted to the left 
 during the whole observations due to the imperfect tracking accuracy.}
\label{fig:mapcrab}
\end{figure}

Figure~\ref{fig:mapcrab} shows the count map around the Crab pulsar. 
This map is composed of $5\times 5$ 1-second exposures,
where the dark counts have been subtracted and the flat-field correction is applied.
The Crab nebula and surrounding stars are successfully imaged (see Appendix \ref{app:sky} 
for imaging a wider field of the sky),
although the Crab pulsar is too faint to be identified in this image.
The count rate of the Crab pulsar can be estimated in two ways:
the one is derived from well known B and V-band flux studied by many authors (e.g., \cite{Per+93}),
and the other is a fit to the count rates - magnitude correlation observed by our system.
In this fit, only two stars HIP26328 and HIP26159 were used and the slope was fixed $M^{-2.5}$,
where $M$ is a magnitude of these stars.
Both estimates yield consistent results of $\sim 100-200$\,counts/s
considering systematic uncertainty in the convolution of the spectrum and the photon detection efficiency.
No filters were used in this work, which may have caused a systematic error for the both estimations.

\subsection{Timing}
The recorded data include light curves with 100-$\mu$s bins for each channel, the light curve of the PPS signal
and the time stamp when the run was started.
We first identified the PPS-tagged light curve bins
which are used not only for calculating absolute UTC time for each time bin
but also for the correction for time shift of the quartz oscillator that controls the 100\,$\mu$s bins onboard the scaler.

\begin{table}[h!]
\tbl{Ephemeris of the Crab pulsar}{
\begin{tabular}{ll}
\hline
Parameter  & value \\
\hline
Right Ascension (J2000)& 05:34:31.97232 \\
Declination (J2000) & +22:00:52.069 \\
Ephemeris & DE200\\
Epoch& 58557.000000062151230\\
F0  (Hz) & 29.621031250700000 \\
F1 (Hz/s)& $-3.6847325\times 10^{-10}$\\
F2 (Hz/s$^2$)& $9.16730648682962\times 10^{-21}$\\
\hline
\end{tabular}
}
\label{tab:ephe}
\end{table}

We used the TEMPO2 package \citep{hob+06} to transform the time of arrival to the solar system barycenter (TDB).
Used timing parameters are listed in table~\ref{tab:ephe} 
and were provided by the Jodrell bank observatory \citep{Lyn+93}\footnote{http://www.jb.man.ac.uk/~pulsar/crab.html}.

\begin{figure}
\centering
\includegraphics[width=8cm]{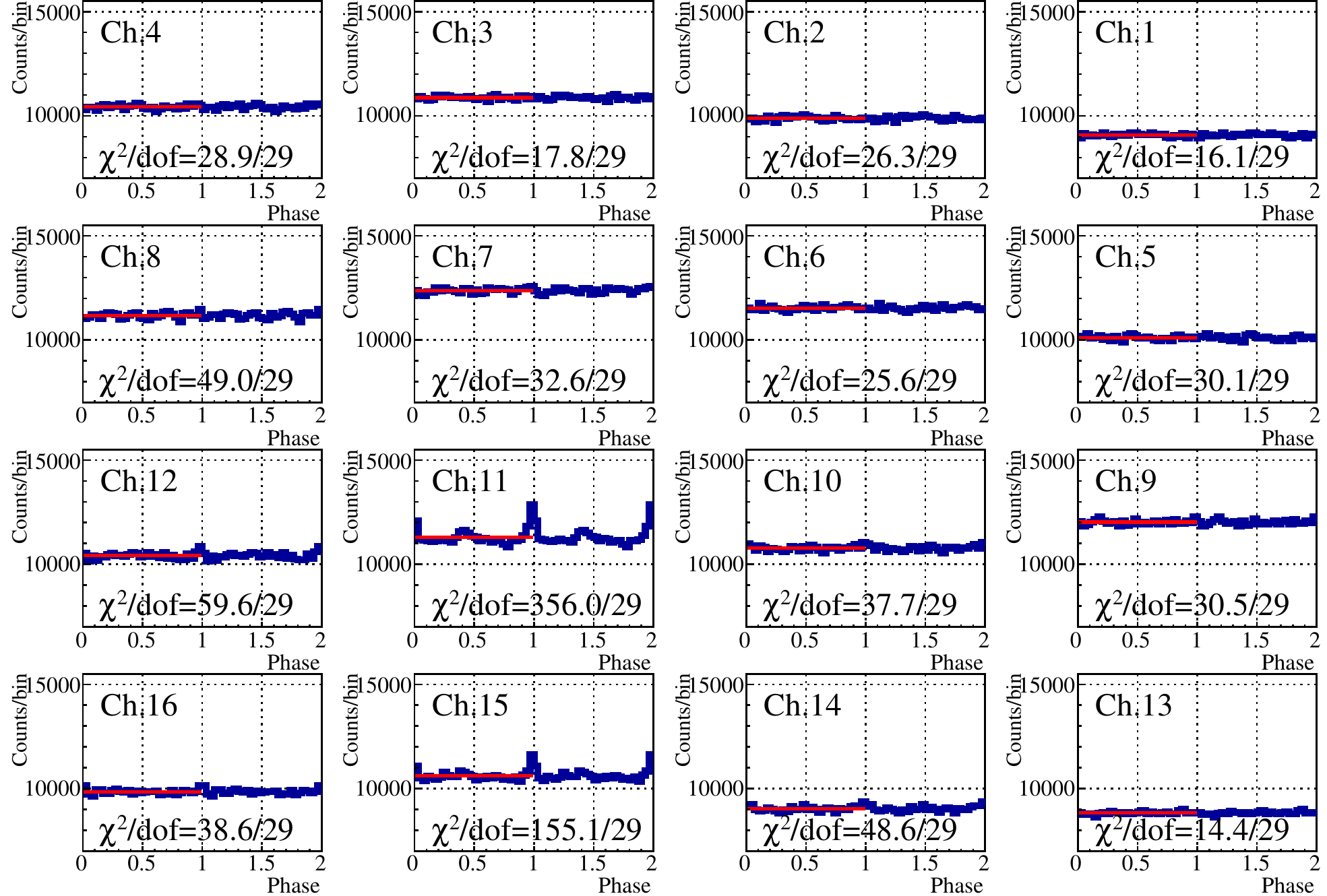}
\caption{Folded light curves for all the pixels of 1-min observation.
Two cycles of the pulsar rotation phase are presented.
The fit results are overlaid and corresponding reduced $\chi^2$ values are also shown.
Each panel is arranged to be the same orientation shown in figure~\ref{fig:mapcrab}. }
\label{fig:lcall}
\end{figure}

Figure~\ref{fig:lcall} shows an example of the folded light curve for the 1-min exposure data (run id 190309211033-1-0),
without the subtraction of the dark counts and the flat-field correction.
The results of a fit to a constant function are overlaid and the corresponding reduced $\chi^2$ are also shown.
When our detection criterion for the periodic signal is set for a confidence level of 0.5\%,
the channels 11, 12, and 15 turned out to contain the Crab pulsar signal in this case.
The same fit was applied to the data after the dark count and flat correction
summed for such pixels with the period detection.

Table\,\ref{tab:results} summarizes the number of pixels detecting the period $n_{\rm pix}$ and 
resulting reduced $\chi^2$ for each run.
It can be seen that in most of runs,
pulses are detected across multiple pixels. 
This is not because the PSF is larger than the pixel size, 
but because of a continuous shift of the image position due to a tracking inaccuracy.
This fact has been directly confirmed by digital camera images of stars
and is also supported by the fact that the pixel 
where the significance of period detection grows shifts from one to another over time, even in a single run.

A phase interval of $0.6-0.8$ was defined as off-pulse 
and then the pulsed component was calculated by subtracting the averaged counts during the off-pulse.
The derived count rates of the pulsed component are also listed in Table\,\ref{tab:results}.
These values are consistent with each other and with
the predicted count rate as mentioned in the previous subsection,
though run id 190309204926-0-0 shows a marginally lower count rate.
This is probably because the Crab pulsar was on the edge of the FoV
and a considerable fraction of the Crab flux could not be detected.
$n_{\rm pix}=1$ supports this interpretation, and the corresponding pixel was channel 3,
 which was indeed located on the outer edge of the GAPD array. 

\begin{table}[h!]
\tbl{Summary of the Crab pulsar observation}{
\begin{tabular}{cccc}
\hline
Run ID  & $n_{\rm pix}$ \footnotemark[$*$] & $\chi^2/{\rm dof}$ \footnotemark[$\dag$]  & $r_{\rm pulse}$ \footnotemark[$\ddag$] \\
\hline
190309194710-0-0 &2& 412.4/99 & $161.5\pm  32.9$ \\
190309200333-0-0 &3& 332.1/99 & $119.1\pm 39.3$ \\
190309203000-0-0 &3& 290.1/99 & $87.3\pm 43.0$  \\
190309204926-0-0 &1& 309.8/99 & $59.1 \pm   22.6$\\
190309204926-1-0 &3& 412.3/99 & $111.4 \pm 23.5$\\
190309211033-1-0 &2& 560.0/99 & $169.7  \pm 32.2$\\
190309212917-1-0 &2& 377.7/99 & $71.1  \pm 35.9$ \\
\hline
\end{tabular}
}
\label{tab:results}
\begin{tabnote}	
	\footnotemark[$*$] Number of pixels detecting the periodic signal.\\
 	\footnotemark[$\dag$] Reduced $\chi^2$ of the fit with a constant.\\
 	\footnotemark[$\ddag$] Count rates of the pulsed component (photons/s) and 68\% statistical error.\\
\end{tabnote}
\end{table}

\begin{figure}
\centering
\includegraphics[width=7cm]{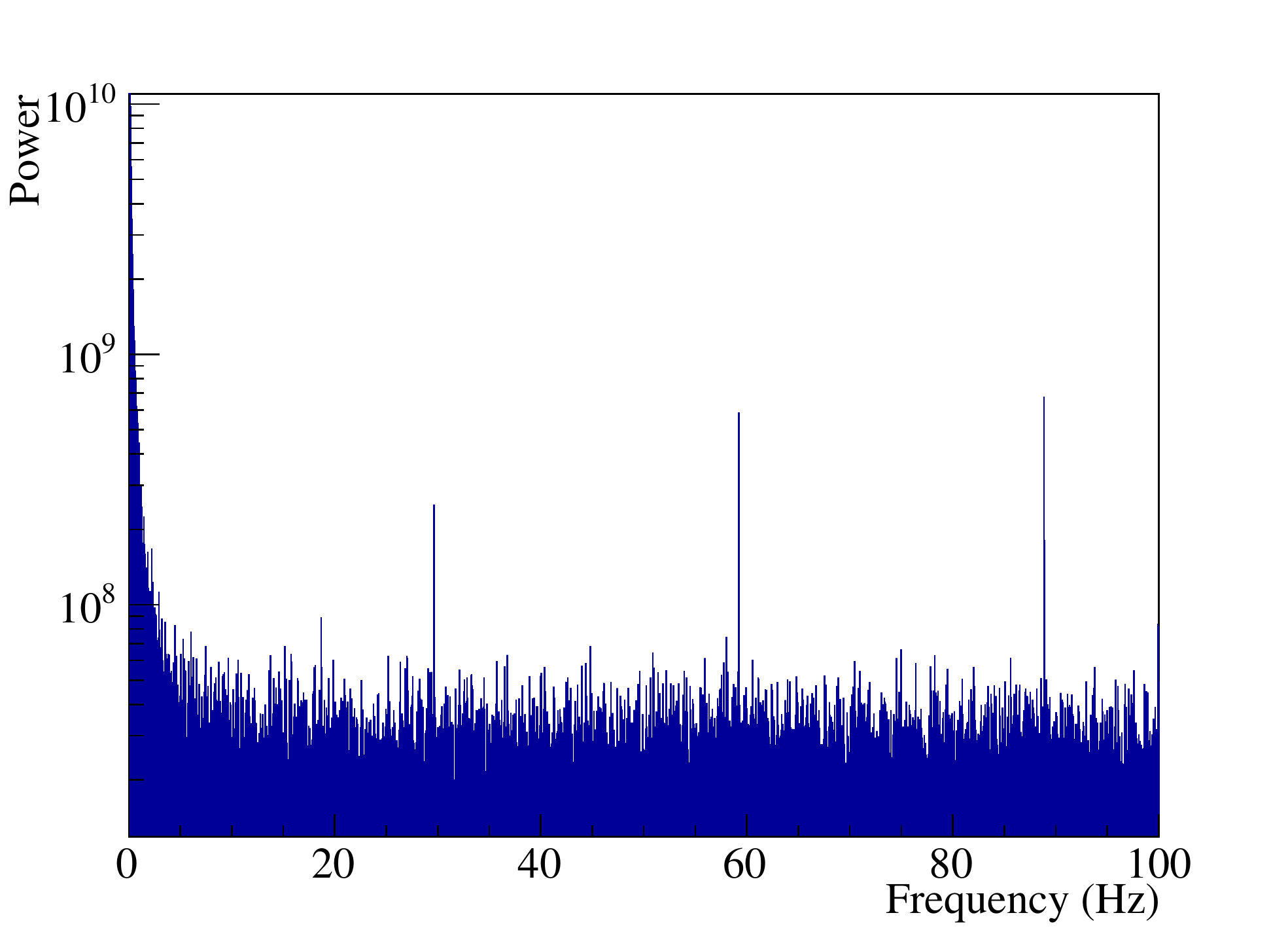}
\caption{Power spectrum derived from all the light curves containing the Crab pulsar.}
\label{fig:crabfft}
\end{figure}

 Power spectra are also calculated by Fast Fourier Transform (FFT)  for all the period-detecting channel data.
% and the clear peaks are commonly found from the pixels exposed to the same sky direction.
Figure~\ref{fig:crabfft} shows the power spectrum obtained from the light curve of all observations where such periodicity was detected.
The normal mode can be identified at $29.6213 \pm 0.0001$\,Hz in TDB time coordinate,
which is consistent with the prediction from the radio ephemeris of the Crab pulsar,
from 29.621208 down to 29.621206\,Hz during the whole acquisition.

\begin{figure}
\centering
\includegraphics[width=8cm]{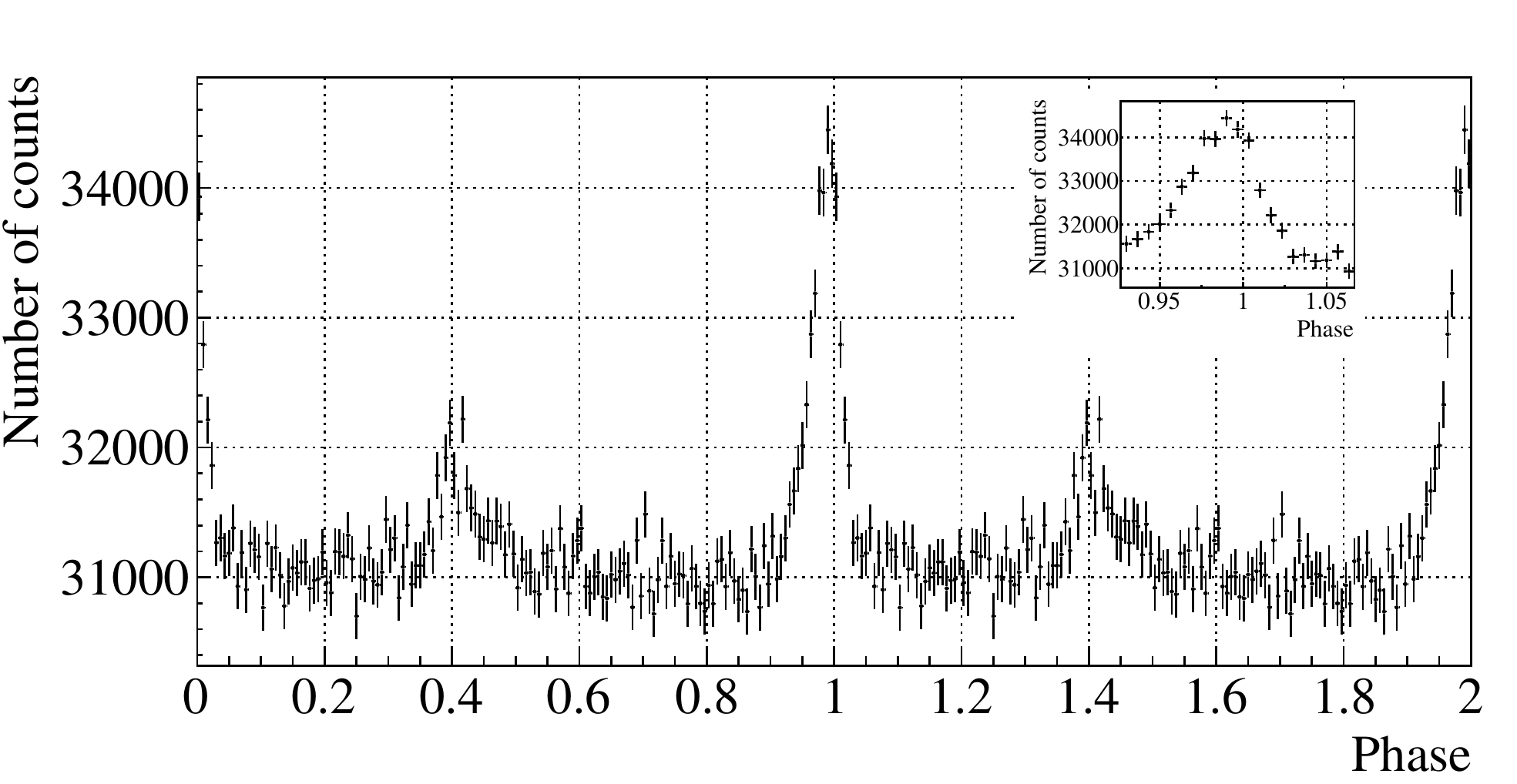}
\caption{Observed pulse profile of the Crab pulsar. 
The width of the time bin is 200\,$\mu$s.
The inset panel shows the close view around the phase=1
which corresponds to the radio main pulse (1.4\,GHz, JBO).}
\label{fig:crabpulse}
\end{figure}

Finally, the integrated pulse profile as shown in figure~\ref{fig:crabpulse} was
 obtained by summing all the period-detecting pixels. 
With the effective exposure of 7\,minutes,
both the main pulse and the interpulse were successfully detected.
That the optical pulse slightly leads the 1.4\,GHz radio pulse
can also be observed.

\section{Discussion and conclusion}
The system developed in this study successfully detected optical pulses of the Crab pulsar
using an amateur telescope under the relatively bright night sky in the city.
Capability of single photon detection and time resolution played a crucial role in this high sensitivity.
Our system was proved to be promising even though further improvements to the equipment are required.
Since the current DAQ system is observing only light curves with a time bin of $100\,\mu$s,
there is a potentially better timing resolution, for example by tagging a time stamp to each photon.
On the other hand, finer time bins require sufficient number of photons in each bin
in order to discuss the time variability.
Although this work was intended to be a demonstration,
only 7\,minutes of data were obtained for the Crab pulsar.
More photon statistics are needed to discuss the variability of the pulsed component
such as the short time scale stability as discussed in \citet{kar07} and 
a few \% enhancements accompanying giant radio pulses \citep{she+03,str+13}.
To realize more efficient observation, longer exposure or mounting the system on a telescope with a larger collection area
is required.

From the viewpoint of photometry, a larger sensor or wider FoV is preferred.
Astronomical objects with fast variability are by nature expected to be compact,
so an expanded FoV is not necessarily required if such target position is well known, like the Crab pulsar.
With larger FoV, however, the system will be tolerant of pointing error, ambiguous tracking accuracy, and scintillation of the object.
In order to achieve an accurate flux measurement, the whole image of the object should be contained in the FoV.
At the same time, larger GAPD area is directly related to an increase of readout channels.
Thus a more integrated DAQ system employing FPGA is under development in our project.

A much larger FoV, for example by a factor of $>10^2$, would enable this system 
to search for transients with uncertain position information such as gamma-ray bursts
 and (non-repeating) fast radio bursts.
Searching for micro meteors from supernova ejecta \citep{ami+20} might be another interesting target.
In addition, observing reference stars in the same FoV could improve the photometric accuracy.

Another possible application could be 
observations of sub-second time-scale occultation of stars by small asteroids \citep{tan+07} and Kuiper belt objects.  
Several projects in the world has already been ahead for such observations
 such as TAOS II \citep{leh+12}, CHIMERA \citep{har+16},  OASES \citep{ari+19}, Tomo-e GOZEN \citep{sako18}
 using fast CMOS cameras. 
 More opportunity could be expected to observe fainter and faster transients 
 since our GAPD array is more sensitive than CMOS.  
 And also portability is advantageous for multipoint observations of asteroidal occultation
 which happen in limited locations.
 Since the system is sufficiently small, for example without powerful cooling equipment, 
our system will achieve portability once the integrated DAQ is developed
which is expected to have low power consumption and mass productivity.

\bigskip
\begin{ack}
This work was supported by JSPS KAKENHI Grant Numbers JP26610054 and JP20H01940
and also by Yamada Science Foundation.
The authors thank Y. Yatsu and H. Okamura for their useful advice in selecting the telescope and the equatorial mount,
S. Gunji, K. Ioka, and F. Yoshida for continuous discussion during the researches. 
People from Small Astronomers' Society, especially K. Inoue, S. Sawa and M. Takahashi are appreciated 
for their support and advice during the installation of the telescope and other instruments.
Nikuni Corporation financially supporting Yamagata Astronomical Observatory is also appreciated.
Finally, we are deeply grateful to the people of Hamamatsu Photonics K. K. for developing and producing the GAPD array.

\end{ack}

\appendix
\section{Pointing and focussing}
\label{app:focus}
\begin{figure}
\centering
\includegraphics[width=8cm]{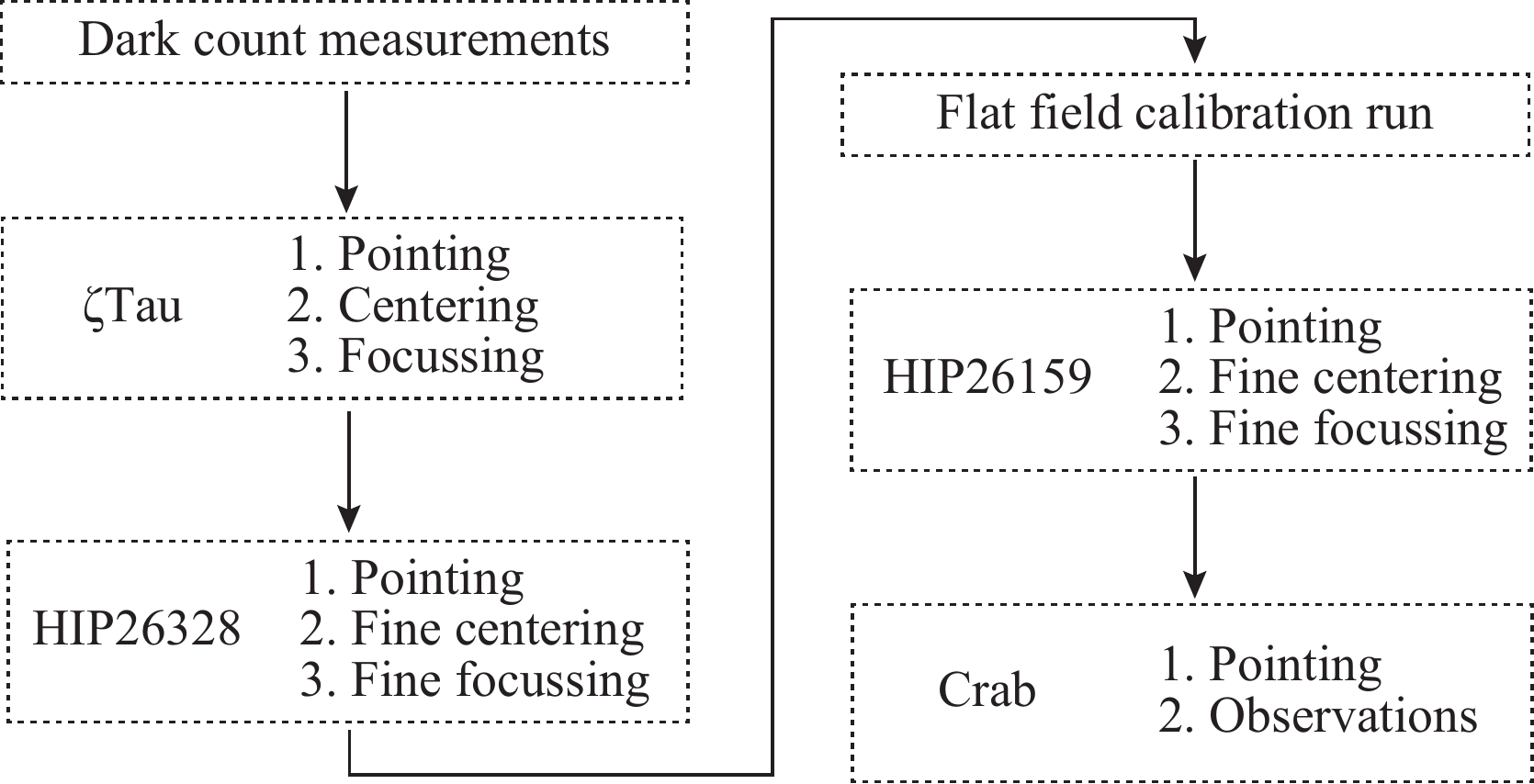}
\caption{Observation sequence.}
\label{fig:seq}
\end{figure}

Focussing procedures are described as follows.
First the telescope was pointed roughly at $\zeta$Tau, a 3rd magnitude star near the Crab pulsar.
Mosaic images of a wider field with a one-second exposure were made 
until a defocused ring-like image of the bright star was found.
Then many trials were conducted to adjust the focus by turning the built-in knob, and checked by mosaic taken each time.
At the same time, the star was moved to the center of the FoV by the fine adjustment of the equatorial mount.
$\zeta$Tau is enough bright to saturate the output signals, when the sensor is close to the focal plane.
Second, the telescope was automatically slewed to HIP26328, which is a 6.88 mag star closer to the Crab pulsar.
By checking the mosaic again,
HIP26328 was turned out to be away from the center of the FoV by $\sim 1$\,arcminutes.
This is probably due to a slight misalignment of the polar axis of the equatorial.
Since there are no other imaging devices are equipped,
the mosaic imaging by the custom MPPC was the only way to confirm the star position
and to allow manual readjustment of the direction of the telescope.
The fine focus was adjusted until the image of the star was successfully contained in a single FoV.
Next the telescope was slewed to HIP 26159, closer to the Crab pulsar than HIP 26328.
Centering and fine focussing were checked.
Indeed, the shift to the center of the image was only several pixels, and further focus adjustment was not necessary.
Finally the telescope pointing was set to the Crab pulsar.
Therefore the Crab pulsar was naturally expected to appear off-center of the FoV.
This is consistent with the position where the pulsation was found as shown in figure~\ref{fig:mapcrab}.
Figure~\ref{fig:seq} summarizes the whole procedure of the observations.

\section{Sky and star observation}
\label{app:sky}

\begin{figure}
\centering
\includegraphics[width=7cm]{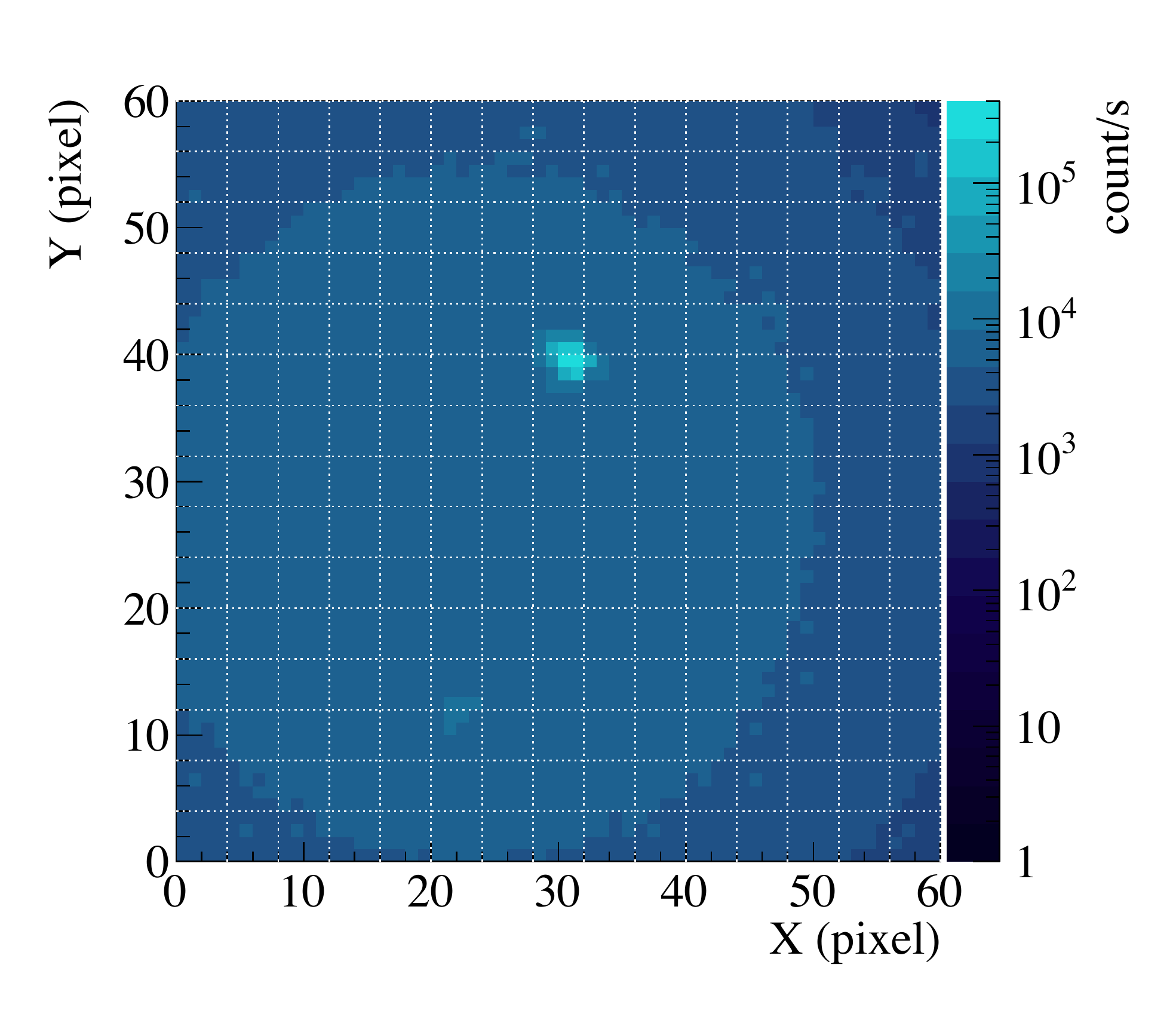}
\caption{Count map around the HIP 26328 region composed of $15\times 15$ observations.
HIP 26328 is located at $(X, Y)=(30,40)$ and a fainter star could be identified around (20,15).}
\label{fig:skymap}
\end{figure}

Figure~\ref{fig:skymap} shows a mosaic count map around the HIP 26328 region,
 in a logarithmic scale.
This map is composed of 1-second $15\times 15$ exposures,
where the dark counts are subtracted and the flat-field correction is applied.
Mechanical vignetting caused by the sensor box is apparent.
A 7-mm diameter hole is located $\sim 35$\,mm above the sensor
and leads to a gradual and radial decrease of the background field. 
This trend is quantitatively confirmed by using ROBAST ray-trace simulator \citep{oku16}.
The map is fit by a disk with a radial gradient in addition to the two-dimensional Gaussian at the star position,
which yields the PSF of $17.1^{\prime\prime}$ (FWHM).
This value indicates the tail of the PSF contaminates the neighboring pixels
and is slightly worse than the star image located at (13, 16) in figure~\ref{fig:mapcrab}.
This value is also worse than the achieved value using a digital camera as mentioned in section \ref{sec:obs},
probably due to manual focus.
To achieve a better PSF and reproductivity, an additional stage in the $Z$-axis is naturally required.

\end{document}